\documentclass[twocolumn,showpacs,preprintnumbers,amsmath,amssymb]{revtex4}
\usepackage{graphicx}
\usepackage{dcolumn}
\usepackage{bm}

\newcommand{\bef}{\begin{figure}}
\newcommand{\eef}{\end{figure}}
\newcommand{\nn}{\nonumber}
\newcommand{\be}{\begin{equation}}
\newcommand{\ee}{\end{equation}}
\newcommand{\bea}{\begin{eqnarray}}
\newcommand{\eea}{\end{eqnarray}}
\newcommand{\bp}{{\bf{p}}}
\newcommand{\bn}{{\bf{n}}}
\newcommand{\la}{\langle}
\newcommand{\ra}{\rangle}
\begin{document}

\title{Two  Photon Correlation in Anisotropic Quark-gluon plasma (aQGP).}

\author{Payal Mohanty, Mahatsab Mandal and  Pradip K Roy}
\email{Payal.Mohanty@saha.ac.in, mahatsab.mandal@saha.ac.in,pradipk.roy@saha.ac.in}
\medskip

\affiliation{HENPP Division, Saha Institute of Nuclear Physics, 1/AF, Bidhannagar, Kolkata 700064, India}

\date{\today}

\begin{abstract}
The only way to obtain the space-time structure of heavy ion collision 
is through the study of two-particle momentum correlations. Thus 
we have studied the intensity correlation for the photons 
at most central collision at RHIC energy having 
fixed transverse momentum of one of the photons ($k_{1T}$= 2 GeV) 
to have an idea about the emission zone in presence of initial momentum space anisotropy. 
For the evolution in aQGP, the {\em free streaming interpolating} model 
with fixed initial condition has been used. Whereas for the space-time evolution 
for both the isotropic Quark Gluon Plasma (iQGP) and the hadronic phases, 
relativistic (1+2)d hydrodynamics model 
with cylindrical symmetry and  boost invariance along longitudinal direction has been used. The variation of  
Bose-Einstein correlation function (BECF), $C_2$, for two identical photons 
as a function of $q_{out}$, $q_{side}$ and $q_{long}$ is evaluated. 
Our analysis is based on  the sensitivity of the momentum space anisotropy 
on the correlation function and also on the extracted HBT radii from it. 
It is observed that the value of $C_2$ as function of $q$'s is reduced 
for anisotropic scenario compared to that of isotropic case for all the directions. 
As a result, the spatial dimension of the reaction zone increases  
due to the presence of momentum space anisotropy in the medium in all directions. 

\end{abstract}

\maketitle


\section{Introduction}
The prime objective of heavy ion collisions (HIC) at relativistic energies is  
to create and  explore the properties of novel state of partonic matter, 
known as Quark Gluon Plasma (QGP). Recently, enormous experimental efforts  at 
Relativistic Heavy Ion Collider (RHIC) and Large Hadron Collider (LHC) 
are carried out in this direction. The method of two particle 
intensity interferometry, commonly known as Hanbury-Brown and Twiss (HBT) 
interferometry~\cite{HBT}, has been used extensively in both theory and 
experiment to obtain the spatial and temporal informations of the particle 
emission zones created in HIC~\cite{pratt,uaw,hb3}. This method was first 
introduced in HIC in the hadronic sector through the study of quantum statistical 
correlation between identical pions which provided valuable inputs for 
the space-time description of the system  at the freeze out surface~\cite{uaw}. 
In contrast to hadrons, the study of two-particle intensity interferometry of 
electromagnetic (EM) radiations, such as photon and dilepton interferometry 
~\cite{peressounko,fmann,AlamHBT,BMSPRL,DKS_HBT,payal_HBT,Rupa_HBT}, 
are more effective as they shed light on the dynamics of the collision 
from the entire evolution.
Owing to large mean free path compared to the size of the system formed in HIC, 
the  EM radiations
travel unscathed from the entire evolution of the fireball without 
further re-scattering  with the surrounding medium and hence can provide 
information of  the history of  evolution of the hot matter
created in HIC ~\cite{larry,ja1,ja2,gale,payal_flow}. 

The hydrodynamical model proposed by Bjorken~\cite{bjorken} takes care 
of the evolution dynamics and is used 
with great triumph to describe the heavy ion collision data which assumes 
the system to be in local thermal equilibrium and isotropy in momentum space.  However, the 
thermalization in the HIC is a debatable issue. Due to poor knowledge of the
isotropization and thermalization time scales ($\tau_{iso}$ and $\tau_{therm}$ respectively),  
one need  not  assume the hydrodynamical 
behavior from the very beginning. There is an additional factor 
that needs to be highlighted here. 
In a realistic scenario, due to the rapid longitudinal expansion at the onset of the QGP phase, 
anisotropy arises in $p_T-p_L$ plane with 
$\la p_L^2 \ra \ll \la p_T^2 \ra $ in the local rest frame.
With time,  such asymmetry dies out with secondary partonic interactions. After which the 
system is considered to be isotropic and thermalized at proper time $\tau_{iso}$ and 
beyond $\tau\geq\tau_{iso}$  the system can be 
treated hydrodynamically. 
To include such momentum anisotropy in pre-equilibrium stage of the QGP phase, a 
simple phenomenological model is adopted from  refs.~\cite{schenke,MMprl100,mauricio}.
In accordance with this model, there are two parameters: plasma momentum space anisotropy 
($\xi$) and the hard momentum scale ($p_{hard}$) which take care of the anisotropic effect. 
We assumed two time scales here; (1) the initial QGP formation time, $\tau_i$, 
and (ii) the isotropization time, $\tau_{iso}$, when the isotropy in momentum space is
achieved and they should fulfill the criteria that $\tau_i \leq \tau_{iso}$. In absence of 
anisotropy, $\tau_i = \tau_{iso}$. 

In the present work, we  study photon interferometry at most central 
RHIC initial conditions at $\sqrt{s_{NN}}$ = 200 GeV including momentum anisotropy 
in the pre-equilibrium QGP phase. We devote our analysis  for full evolution of the fireball to observe
the anisotropic effect  on the size of the emission zone. 
Such an analysis helps in understanding the effect of anisotropy on 
the space-time dynamics of the evolving source. 
The free streaming interpolating model
with fixed initial condition has been used for evolution in aQGP and (1+2)d hydrodynamical model 
with cylindrical symmetry and longitudinal boost invariance is used for $\tau\ge\tau_{iso}$. 
Here we have relaxed the assumption of local isotropy and 
considered the plasma with local anisotropy in $p_T-p_L$ plane in 
pre-equilibrium aQGP phase. 
We have assumed the following two sets of initial conditions for the analysis;  
SET-I: $\tau_i$=0.147 fm/c, $T_i$= 446 MeV and SET-II: $\tau_i$=0.24 fm/c, $T_i$= 350 MeV. 
In the present approach, we adopt two-stage evolution scenario. For aQGP evolution we use $\tau_i$ 
and $T_i$ as the initial conditions. However, for hydrodynamic evolution, i.e, iQGP evolution $\tau_{iso}$ 
and $p_{hard}(\tau_{iso})$ can be considered as the initial conditions.
 The thermalized isotropic QGP is assumed with sufficiently high energy densities 
 at  $\tau=\tau_{iso}$.  Afterwards, with  expansion,  the energy
density reduces,  hadronization begins at $\tau_q$ (pure QGP phase ends here).  
The system then undergoes a phase transition at transition temperature ($T_c$) and 
transforms to a hadronic gas phase at $\tau_h$. With further expansion, 
the energy density reduces further and finally reaches freeze-out at  $T_f$, 
called freeze-out temperature at freeze-out time ($\tau_f$). 
 In this work, we shall study the sensitivity of the two-photon correlation functions 
on the isotropization time ($\tau_{iso}$), whereas the authors in  \cite{Rupa_HBT} have investigated 
the behaviour of correlation functions by varying the formation time ($\tau_i$).

 The article is organized as follows. We have started with 
 the definition and formulation of Bose-Einstein correlation function (BECF) in 
 Sec.~\ref{sec_defn}. The thermal emission rate of  photons used for the present 
 calculation is discussed in Sec.~\ref{sec_rate}. The model used for space - time 
 evolution is briefly outlined in Sec.~\ref{sec_evo}. In Sec.~\ref{sec_result} we 
 discuss the sensitivity of our results to the initial momentum anisotropy with varying 
 $\tau_{iso}$. Finally we  summarize  in Sec.~\ref{sec_summ}. 
\section{Definition and Formalism}
\label{sec_defn}
The Bose-Einstein correlation function (BECF) for two photons 
with momenta $\vec{k_1}$ and $\vec{k_2}$ is defined as,
\begin{equation}
C_{2}(\vec{k_{1}},\vec{k_{2}})=\frac{P_{2}(\vec{k_{1}},\vec{k_{2}})}
{P_{1}(\vec{k_{1}}) P_{1}(\vec{k_{2}})} 
\label{eq_c2}
\end{equation}
where
\begin{equation}
P_{1}(\vec{k}) = \int d^{4}x~\omega(x,k);~~~\omega(x,k)=E\frac{dR}{d^3k}
\label{eq_P1}
\end{equation}
and
\bea
&&P_{2}(\vec{k_{1}},\vec{k_{2}})= 
P_{1}(\vec{k_{1}})P_{1}(\vec{k_{2}})\nn\\
&&+\frac{1}{2}\int d^{4}x_{1} d^{4}x_{2} ~\omega (x_{1},K) 
\omega (x_{2},K)~\cos(\Delta x^{\mu} \Delta k_{\mu})\nn\\ 
\label{eq_P2}
\eea
where $\vec{k_i}=(k_{iT}\cos \psi_i, k_{iT}\sin \psi_i,k_{iT}\sinh y_i)$ 
is the three momentum of the two identical photons with $i=1,2$,  
$K=(k_1+k_2)/2$ is the average  momentum, 
$\Delta k_\mu=k_{1\mu}-k_{2\mu}=q_\mu$,
$x_i$ and $k_i$ are the four co-ordinates for position and momentum 
variables respectively and
$\psi_i$'s are the angles made by $k_{iT}$ with the x-axis of each photon. 
The inclusion of the spin of the  real photon  
will reduce the value of $C_2-1$ by 1/2. 

We shall be presenting the results as  functions of 
outward ($q_{out}$), side-ward ($q_{side})$ and longitudinal ($q_{long}$) 
momenta which can be expressed in terms of the transverse momentum of 
individual pair as follows~\cite{uaw};
\begin{eqnarray}
q_{side}&&=\left|\vec{q_T}-q_{out}\frac{\vec{K_T}}{K_T}\right|\nn\\
&&=\frac{2k_{1T}k_{2T}\sqrt{1-\cos^2(\psi_1-\psi_2)}}
{\sqrt{k_{1T}^2+k_{2T}^2+2k_{1T}k_{2T}\cos(\psi_1-\psi_2)}}\nn\\   
\eea
\bea
q_{out}&&=\frac{\vec{q_T}.\vec{K_T}}{|K_T|}\nn\\
&&=\frac{(k_{1T}^2-k_{2T}^2)}{\sqrt{k_{1T}^2+k_{2T}^2+2k_{1T}k_{2T}\cos(\psi_1-\psi_2)}}\nonumber\\
\eea
\bea
q_{long}=k_{1z}-k_{2z}=k_{1T}\sinh y_1-k_{2T}\sinh y_2
\label{eq_q}
\end{eqnarray}
where $k_{iT}$ is the individual transverse momentum and $y_i$ is the rapidity. 
It may be mentioned that the BEC function has values 
$1\leq C_{2}(\vec{k_{1}},\vec{k_{2}}) \leq 2$ 
for a chaotic source. These bounds are from
quantum statistics. 
The source dimensions can be obtained
by parameterizing the calculated correlation function with the
empirical Gaussian form~\cite{pratt};
\begin{eqnarray}
&&C_2(q,K)  =1+\lambda \exp(-R^2_iq^2_i) \nn\\
& = &1+\lambda \exp(- R_{side}^2q_{side}^2-R_{out}^2q_{out}^2-R_{long}^2q_{long}^2)\nn\\
\label{eq_prmt}
\end{eqnarray}
where $i$ stands for side, out and long. Thus 
$R_{side}$, $R_{out}$ and $R_{long}$ appearing in Eq.~\ref{eq_prmt},  
are commonly referred to as “HBT radii”, which are measures of 
Gaussian widths of the source size.  
The deviation of $\lambda$ from
1/2 will indicate the presence of non-thermal sources, 
while the radius, $R_{side}$ corresponding to $q_{side}$
 is closely related to the transverse size of the system. 
The radius, $R_{out}$ corresponding to $q_{out}$  measures both the
transverse size and the duration of particle emission and $R_{long}$ corresponding 
to $q_{long}$ is the measure of longitudinal dimension of the system ~\cite{uaw,hb3,rischke,hermm,chappm}. 
\section{Emission rate of  photons }
\label{sec_rate}
\subsection{Photon emission rate  in anisotropic QGP}
\label{sec_rate_aniso}
In the present work, the QCD annihilation 
($q\bar{q}\rightarrow g\gamma$) and 
Compton ($q(\bar{q})g\rightarrow q(\bar{q})\gamma$) 
processes contribute to the photon spectra from QGP phase  
which  has been calculated considering the quarks to be 
 massive to avoid the divergence. 
The source function is related to the thermal emission rate 
of  photons  per unit four volume which is given by~\cite{LB,schenke}:
\bea
E\frac{dR}{d^3k}&=& \frac{\mathcal{N}}{2(2\pi)^3}
\int \frac{d^3p_1}{2E_1(2\pi)^3}
\frac{d^3p_2}{2E_2(2\pi)^3}\frac{d^3p_3}{2E_3(2\pi)^3}\nn\\
&&\times(2\pi)^4\delta^{(4)}(p_1+p_2-p_3-k)\overline{\lvert\mathcal{M}\rvert ^2}\nn\\
&&\times f_1(\bp_1, p_{hard}, \xi) f_2 (\bp_2, p_{hard}, \xi)\nn\\
&&\times[1\pm f_3 (\bp_3, p_{hard}, \xi)]\nn\\
\label{eq_om}
\eea
where ${\mathcal N}$ is the over all degeneracy for the reactions 
under consideration, $\overline{\lvert\mathcal{M}\rvert ^2}$ is 
the square of the invariant amplitude for the processes~\cite{wong} under consideration 
(here $q\bar{q}\rightarrow g\gamma$ and $qg \rightarrow q\gamma$), $f_i$'s  
are the anisotropic distribution functions of the constituent partons in the medium.
 
 In this work, we have assumed  a system with a high momentum-space anisotropy 
 where particles move in the specific direction. 
 In such a scenario, the phase space distribution function can be obtained 
 by compressing or stretching an arbitrary distribution along one 
 direction in momentum space and can be expressed as follows~\cite{RomSland}:
\be
f(\bp,\xi,p_{hard})=f_{iso}
\left(\sqrt{\bp^2+\xi(\bp\cdot\bn)^2},p_{hard}\right)
\label{eq_f}
\ee
where $\bn$ is direction of anisotropy. As mentioned earlier,  
$\xi$   is a parameter of momentum space anisotropy and  $p_{hard}$ is 
the hard momentum scale  which is directly related to the average 
 momentum of the partons. $p_{hard}$ has a direct relevance with the temperature (T) of 
 medium in isotropic scenario. We further assume that $f_{iso}$ is the Fermi-Dirac (Bose-Einstein) 
 distribution function for quarks (gluons).  The anisotropy parameter $\xi$ is related 
 to transverse momentum ($p_T$) and longitudinal momentum ($p_L$)  of the constituents 
 via the following relation:
\be
\xi=\frac{\la p_T^2\ra}{2\la p_L^2\ra}-1
\label{eq_xi}
\ee
When $\xi=0$, the system is locally isotropic, but that 
does not imply the system to be in local thermal 
equilibrium unless  $f_{iso}$ is an equilibrium 
distribution function.
\subsection{Photon emission rate from thermal medium}
\label{sec_rate_thermal}
Beyond $\tau\geq\tau_{iso}$, thermal photons emerge just after the system thermalizes 
from both iQGP due to partonic interactions and from hot hadronic matter due 
to interactions among the hadrons.
The rate of thermal photon production per unit space-time volume 
is given by~\cite{larry,gale,weldon} (see ~\cite{ja1} for a review):
\begin{equation}
E\frac{dR}{d^3k}=\frac{g^{\mu\nu}}{(2\pi)^3}\,{\mathrm Im}\Pi^R_{\mu\nu}
f(E,T)
\label{thermalphotonrate}
\end{equation}
where ${\mathrm Im}\Pi_{\mu}^{\mu}$ is the
imaginary part of the retarded photon self energy and $f(E,T)$ is the
thermal phase space distribution. 
For an expanding system, the energy $E$ should be replaced by
$u_\mu k^\mu$, where $k^\mu$ and $u^\mu$ are the four
momentum and the fluid four velocity respectively. 

\subsubsection{Photon emission rate in isotropic QGP}
\par
 Thermal photons are produced from isotropic QGP with the interactions of the
thermal quarks and gluons via QCD Compton and annihilation processes.  
To calculate the imaginary part of the photon self energy appearing in 
the Eq.~\ref{thermalphotonrate} Hard Thermal Loops~\cite{braaten} approximation 
has been used. The complete calculation of the emission rate of 
photons from QGP to order O($\alpha\alpha_s$)has been done by 
resuming the ladder diagrams in the effective theory ~\cite{arnold}. 
\subsubsection{Photon emission rate in hadronic phase}
\par
 A set of  hadronic reactions
with all possible isospin combinations has been considered for the 
production of
photons~\cite{npa1,npa2,turbide} from the hadronic matter.
The effect of hadronic dipole
form factors has been taken into account in the present
work as in~\cite{turbide}.

\section{Space-time evolution}
\label{sec_evo}
For a dynamically evolving system the total photon yield can be evaluated 
by convoluting the static thermal emission rate (discussed in the previous section) 
with the expansion dynamics. In the present work the space time evolution  is 
described as follows~\cite{LB}. The system evolves anisotropically from $\tau_i$ to 
$\tau_{iso}$, where one needs to know the time dependence of 
$p_{hard}(\tau)$ and $\xi(\tau)$. We follow the work of ref~\cite{MMprl100} for
the evolution in aQGP phase. As the radial flow is not developed properly 
in the initial stage of the collision,  its effect is neglected in the anisotropic 
phase. For $\tau\geq\tau_{iso}$, the system is described by (1+2)d ideal 
hydrodynamics model with cylindrical symmetry~\cite{hvg} and boost invariance 
along the longitudinal direction~\cite{bjorken}. Thus, the entire evolution is 
categorized as follows:
\begin{itemize}
 \item 
When $\tau_i\leq\tau\leq\tau_{iso}$, the 
system evolves anisotropically and  $p_{hard}(\tau)$ and $\xi(\tau)$ are the 
two time dependent parameters. 
\item
When $\tau_{iso}\leq\tau\leq\tau_f$, the system becomes thermalized and evolves 
hydrodynamically with energy density ($\cal E$) and velocity as a function of space and time.
\end{itemize}
Hence, $\tau_{iso}$ is treated as free parameter in the calculation which controls 
the transition  to the hydrodynamic scenario. However, it should be noted that 
by comparing with the data, it is possible to extract $\tau_{iso}$ as is done 
in case of single photon spectra~\cite{LB} and nuclear modification factor 
for the light hadrons~\cite{MM}.

Thus, the one- and two- particle inclusive spectra  can be presented as follows,
\bea
P_1(k)&=&P_1^{aniso}(k)+P_1^{hydro}(k)\nn\\
P_2(k_1,k_2)&=&P_2^{aniso}(k_1,k_2)+P_2^{hydro}(k_1,k_2)\nn\\
&&
\eea
 $P_i^{aniso}$ and $P_i^{hydro}$ can be evaluated using  Eqs.~\ref{eq_P1} and ~\ref{eq_P2} with the help of 
space-time prescription for anisotropic and hydrodynamic scenario  given in the following sections. 
Finally using Eq.~\ref{eq_c2} we obtain  $C_2$ for the full evolution as well as for the individual phases.

In anisotropic prescription,  $P_i$'s from an expanding system 
can be calculated by convoluting the static thermal emission rate ($\omega=E dR/d^3k$) 
with the expansion dynamics which depends on energy density, ${\cal E}(p_{hard},\xi)$.
Using Eq.~\ref{eq_f}
the parton energy density in an anisotropic plasma 
can be factorized in the following manner:
\be
 {\cal E}_0(p_{hard},\xi)
=\int \frac{d^3p}{(2\pi)^3}p^0 f(\bp,\xi)
={\cal E}_{iso}(p_{hard})R(\xi)
\label{eq_epsilon}
\ee
where
${\mathcal R}(\xi)=[1/(\xi+1)+\tan^{-1}{\sqrt{\xi}}/\sqrt{\xi}]/2$ 
and  ${\cal E}_{iso}(p_{hard})$ is obtained by integrating the 
parton distribution functions (Eq.~\ref{eq_f}) for $\xi=0$.

Now let us consider the space time evolution model for the aQGP to know the time 
dependence of $\xi$ and $p_{hard}$.  For this, we follow
the works of Refs.~\cite{MMprl100,mauricio,LB}.  According to this model there can be 
 three possible scenarios : (i) $\tau_{iso}=\tau_i$, the system evolves hydrodynamically 
 so that $\xi = 0$ and $p_{hard}$ can be identified with the temperature (T) of the system, 
 (ii)$\tau_{iso}\rightarrow \infty$, the system never comes to equilibrium, 
 (iii)$\tau_{iso}\geq\tau_i$ and $\tau_{iso}$ is finite, one should  devise a time evolution 
 model for $\xi$ and $p_{hard}$ which smoothly interpolates between pre-equilibrium anisotropy 
 and hydrodynamics. This model can be executed mathematically by generalizing the anisotropic parameter 
 $\xi (\tau)$ as follows;
 
 \begin{equation}
  \xi(\tau) = \left(\frac{\tau}{\tau_i}\right)^{\delta}-1,
  \end{equation}
  The limits $\delta = 0$ and $\xi = 0$ correspond to scenario (i) where 
 expansion is hydrodynamical (thus iQGP corresponds to  $\delta=0$). On the otherhand, the 
 limits  $\delta\neq0$ and  $\xi\neq0$ correspond to scenario (ii) where the system is highly anisotropic 
 (thus aQGP corresponds to  $\delta\neq0$). For the present work we shall be following  scenario (iii) 
  which interpolates between hydrodynamic evolution and the 
 anisotropic evolution. For constructing such an interpolating model, a transition width $\gamma^{-1}$ 
is introduced to take into account the smooth transition from  $\delta\neq0$ to $\delta=0$ at $\tau=\tau_{iso}$
(see Ref.~\cite{mauricio} for details). The time dependence of various quantities are,
therefore, obtained in terms of a smeared step function ~\cite{MMprl100}; 
\begin{equation} 
\lambda(\tau)=\frac{1}{2}(\tanh[\gamma(\tau-\tau_{iso})/\tau_i]+1) 
\label{eq_lamda}
\end{equation}
It is clear from the above equation that for $\tau\ll\tau_{iso}$, $\lambda\rightarrow0$ (anisotropic evolution) 
and for $\tau\gg\tau_{iso}$, $\lambda\rightarrow1$ (hydrodynamic evolution). 

With this, the time dependence of relevant quantities are as follows~\cite{MMprl100,mauricio}:
\begin{eqnarray}
{\cal E}(\tau)&=&{\cal E}_0\,[{\cal U(\tau)}/{\cal U}(\tau_i)]^{4/3},\nn\\
p_{hard}(\tau)&=&T_i\,\left[{\cal U(\tau)}/{\cal U}(\tau_i)\right]^{1/3},\nn\\
\xi(\tau) &=& a^{\delta[1-\lambda(\tau)]}-1,
\label{eq_xirho}
\end{eqnarray}
where,
${\mathcal U}(\tau)\equiv\left[{\mathcal R}
\left(a_{iso}^\delta-1\right)\right]^{3\lambda(\tau)/4}
\left(a_{iso}/a\right)^{1-\delta[1-\lambda(\tau)]/2}$, 
$a\equiv \tau/\tau_i$ and $a_{iso}\equiv\tau_{iso}/\tau_i$. 
The power of ${\mathcal R}$ in ${\mathcal{U}}$ keeps energy 
density continuous at $\tau=\tau_{iso}$ for all $\gamma$. 
In the present work, we have used a {\em free streaming interpolating} model 
that interpolates between early-time longitudinal free streaming and late-time  
ideal hydrodynamic expansion by choosing $\delta=2$.

As mentioned earlier, in hydrodynamic prescription,  $P_i$'s from an expanding system 
can be calculated by convoluting the static thermal emission rate ($\omega=E dR/d^3k$) 
with the expansion dynamics which depends on energy density (${\cal E}(r,\tau)$) and 
radial velocity ($v_r(r,\tau)$).
\bea
&&P_1^{hydro}(k)=\sum_{i}{\int_{i}{\omega_i(E^*,T)d^4x}}\nn\\
&&P_2^{hydro}(k_1,k_2)= P_1^{hydro}(k_1)P_1^{hydro}(k_2)+\nn\\
&&\sum_{i}\frac{1}{2}\int d^{4}x_{1} d^{4}x_{2} ~\omega_i (E^*,T) 
\omega_i (E^*,T)~\cos(\Delta x^{\mu} \Delta k_{\mu})\nn\\
&&
\label{p1p2_hydr}
\eea
The energy, $E^*$  appearing in the thermal phase space factor, $f(E^*,T)$ 
(see Eq.~\ref{thermalphotonrate}) should be replaced by $k^\mu u_\mu$
for a system expanding with space-time dependent  four velocity ($u^\mu$). 
Assuming cylindrical symmetry and longitudinal boost
invariance  $k^\mu u_\mu$ can be expressed as
\be
k^{\mu} u_{\mu}=\gamma_r(k_T\cosh(y-\eta)-v_rk_T\cos\phi),
 \ee
where $v_r(\tau,r)$ is the radial velocity, 
$\gamma_r(\tau,r)=(1-v_r(\tau,r))^{-1/2}$. 

The system produced in  QGP phase reverts back  to  hot hadronic phase  
at a temperature $T\sim T_c$. Thermal equilibrium may be maintained in the hadronic phase 
until the mean free path remains comparable to the system size.  
In Eq.(\ref{p1p2_hydr}), 
$i$ stands for qgp (Q), 
mixed phase (M) (in a 1st order phase transition scenario) 
and hadronic phase (H). 
Thus $P_i$'s for full hydrodynamic evolution can be obtained by summing the 
contribution from individual phase, where the contribution from each
phase can be obtained by choosing the phase space appropriately.

The initial conditions are essential for ideal hydrodynamics and can be  obtained by the following prescription~\cite{LB}. 
\begin{eqnarray}
 T_i^{hydro}&=&p_{hard}(\tau_{iso})\nn\\
\tau_i^{hydro}&=&\tau_{iso}
\end{eqnarray}

The initial conditions are given through the energy density and velocity profile,
\begin{eqnarray}
&&{\cal E}(\tau_i,r)=\frac{{\cal E}_0\,(\xi=0, T_i^{hydro})}{1+\exp(\frac{r-R_A}{\sigma×})}\nonumber\\
&&v_r(\tau_i,r)=v_0\left(1-\frac{{\cal E}_0\,(\xi=0, T_i^{hydro})}{1+\exp(\frac{r-R_A}{\sigma×})}\right) 
\label{initial}
\end{eqnarray}
where ${\cal E}_0\,(\xi=0, T_i^{hydro})$ is the initial energy density which is related to initial 
temperature ($T_i^{hydro}$). Here   $R_A$    
is the nuclear radius and $\sigma$ is the diffusion parameter and taken as 0.5 fm. 
We have taken the transition temperature $T_c$= 175 MeV and fixed the freeze-out temperature $T_f$ =120 MeV. 
 For the QGP and the hadronic phases lattice QCD EoS ~\cite{MILC} (for $T>T_c$) and 
hadronic resonance gas EoS~\cite{bmandja} (for $T<T_c$) have been used respectively.
For the transition region we have used the following parameterization~\cite{entropy_prmt};
\begin{eqnarray}
s(T)=s_q(T) f_q(T)+[1-f_q(T)]s_h(T), \nonumber\\
f_q(T)=\frac{1}{2}(1+\tanh\frac{T-T_c}{\Gamma})
\label{entro}
\end{eqnarray}
where $\Gamma$ is the width parameter and assumes a finite value for the
crosssover transition and for the first order transition this value can be tuned to zero. 
Here the width parameter is taken to be  $\Gamma$=25 MeV.

\section{Results}
\label{sec_result}
With all these ingredients discussed in the  previous sections 
we have evaluated the two-photon correlations as a function of 
$q_{out}$, $q_{side}$ and $q_{long}$ for two sets of RHIC 
initial conditions. In both the cases for $\tau_{iso}>\tau_i$, we have observed a  reduction of $C_2$ 
for anisotropic scenario compared to that of in isotropic case 
by incorporating the initial momentum anisotropy. 
We choose $\tau_{iso}$ in such a way that one of the values corresponds to the isotropic situation ($\tau_{iso}=\tau_{i}$) 
and others corresponds to anisotropic scenario ($\tau_{iso}>\tau_{i}$). 
 So basically we have attempted to examine the sensitivity of momentum anisotropy on $C_2$ 
 by controlling the variable, $\tau_{iso}$. The corresponding  HBT radii can be extracted with the help of 
the parametrization expressed in Eq.~\ref{eq_prmt} and can be compared with the data to extract $\tau_{iso}$. 
\subsection{$C_2$ as function of $q_{out}$} 
\bef[!htbp]
\begin{center}
\includegraphics[scale=0.3]{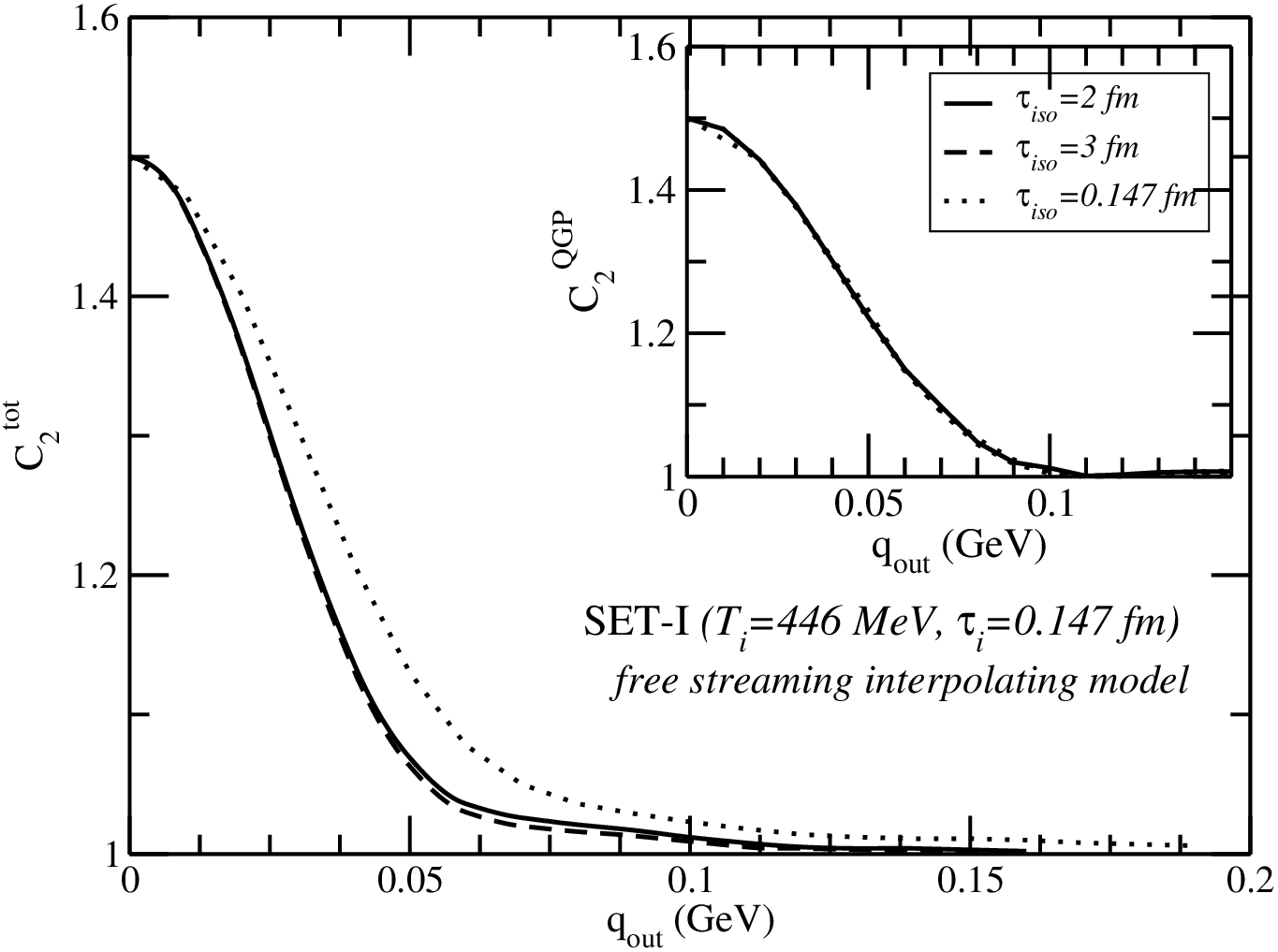}
\caption{Correlation function for photon pairs as a function of 
$q_{out}$  for SET-I ($T_i=446$ MeV and $\tau_i=0.147$ fm/c) 
is plotted with different $\tau_{iso}$ and 
the inset figure is same for QGP (aQGP+iQGP) phase only.}
\label{fig_c2_qout_446}
\end{center}
\eef
\bef
\begin{center}
\includegraphics[scale=0.3]{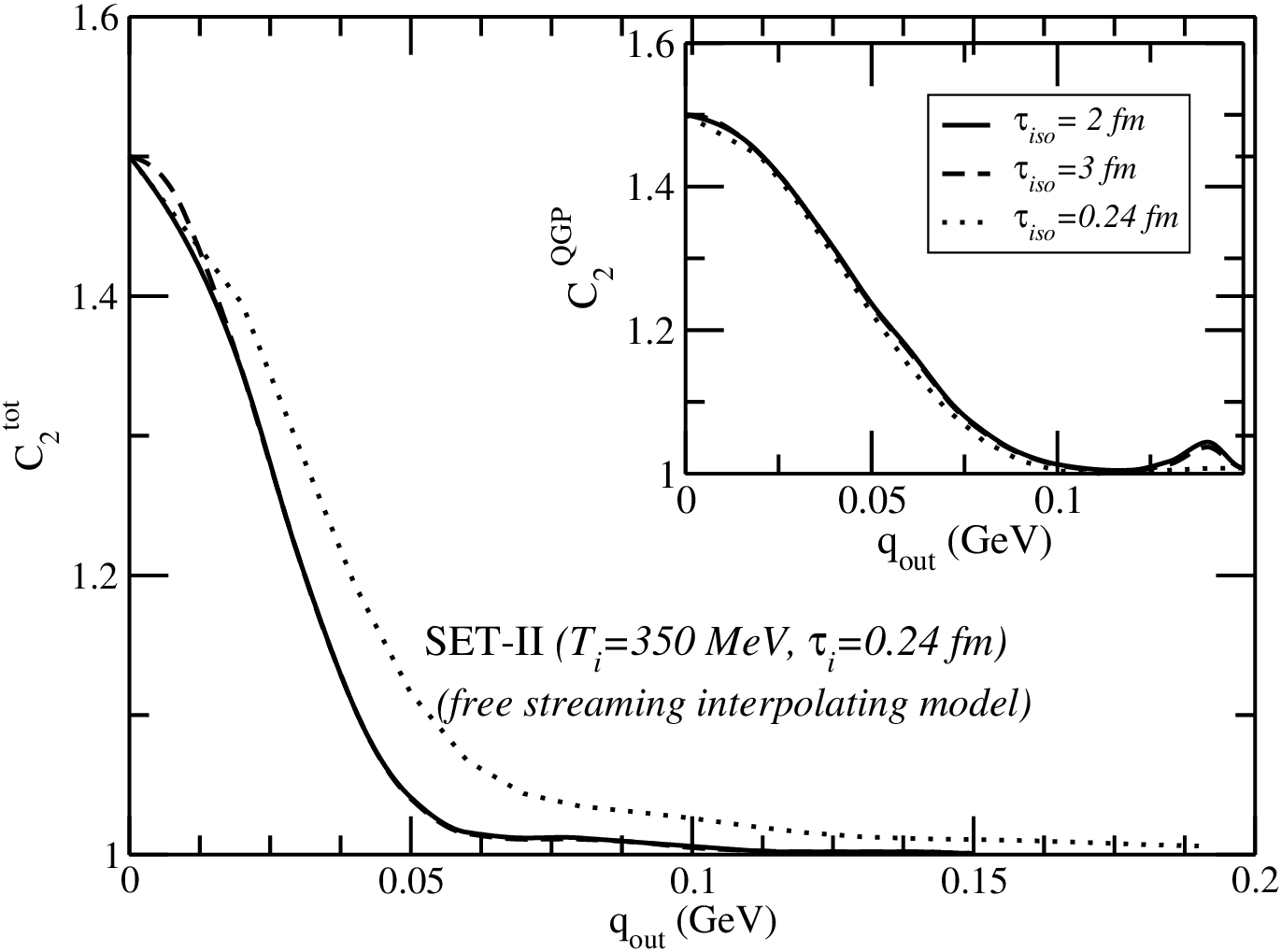}
\caption{Correlation function for photon pairs as a function 
of $q_{out}$  for SET-II ($T_i=350$ MeV and $\tau_i=0.24$ fm/c) 
is plotted with different $\tau_{iso}$ and 
the inset figure is same for QGP (aQGP+iQGP) phase only.}
\label{fig_c2_qout_350}
\end{center}
\eef

By  taking $\psi_1=\psi_2$=0, $y_1=y_2$=0 and fixing the 
transverse momentum of one photon ($k_{1T}$ = 2 GeV) and 
varying the other ($k_{2T}$), we obtain $C_2$ as a function of  $q_{out}$. 
 In Figs.~\ref{fig_c2_qout_446} and \ref{fig_c2_qout_350}, we have plotted the 
variation of $C_2$ as a function of $q_{out}$ in full evolution scenario with  
SET-I and SET-II initial conditions respectively.  
From both the figures, we infer that varying  $\tau_{iso}$, a considerable 
 shift is observed in $C_2$.  
With increasing  $\tau_{iso}$, 
the value of $R_{out}$ (see table~\ref{tb_radii}) which corresponds to $q_{out}$ increases. 
This happens because by increasing $\tau_{iso}$, the system expands slower to 
achieve thermalization and isotropization. 
Whereas in the inset  of 
Figs.\ref{fig_c2_qout_446} and \ref{fig_c2_qout_350} which describe  $C_2$ for the QGP (aQGP+isotropic QGP) phase,
 substantial change is not observed unlike the case for $C_2^{tot}$. 
 This happens because the  flow is not developed 
in early  QGP phase and with the progress of time the thermal energy is transformed into flow energy in later 
stage of the collision, so flow is fully developed in the hadronic stage 
~\cite{payal_flow,payal_HBT} resulting in reduction of $R_{out}$ by increasing $\tau_{iso}$.  
The reduction is mostly affected due to the radial flow as well as the $\tau_{iso}$ 
dependent initial conditions for hydrodynamic evolution which is due to the inclusion 
of momentum space anisotropy. 
\subsection{$C_2$ as function of $q_{side}$} 
\bef[!htbp]
\begin{center}
\includegraphics[scale=0.3]{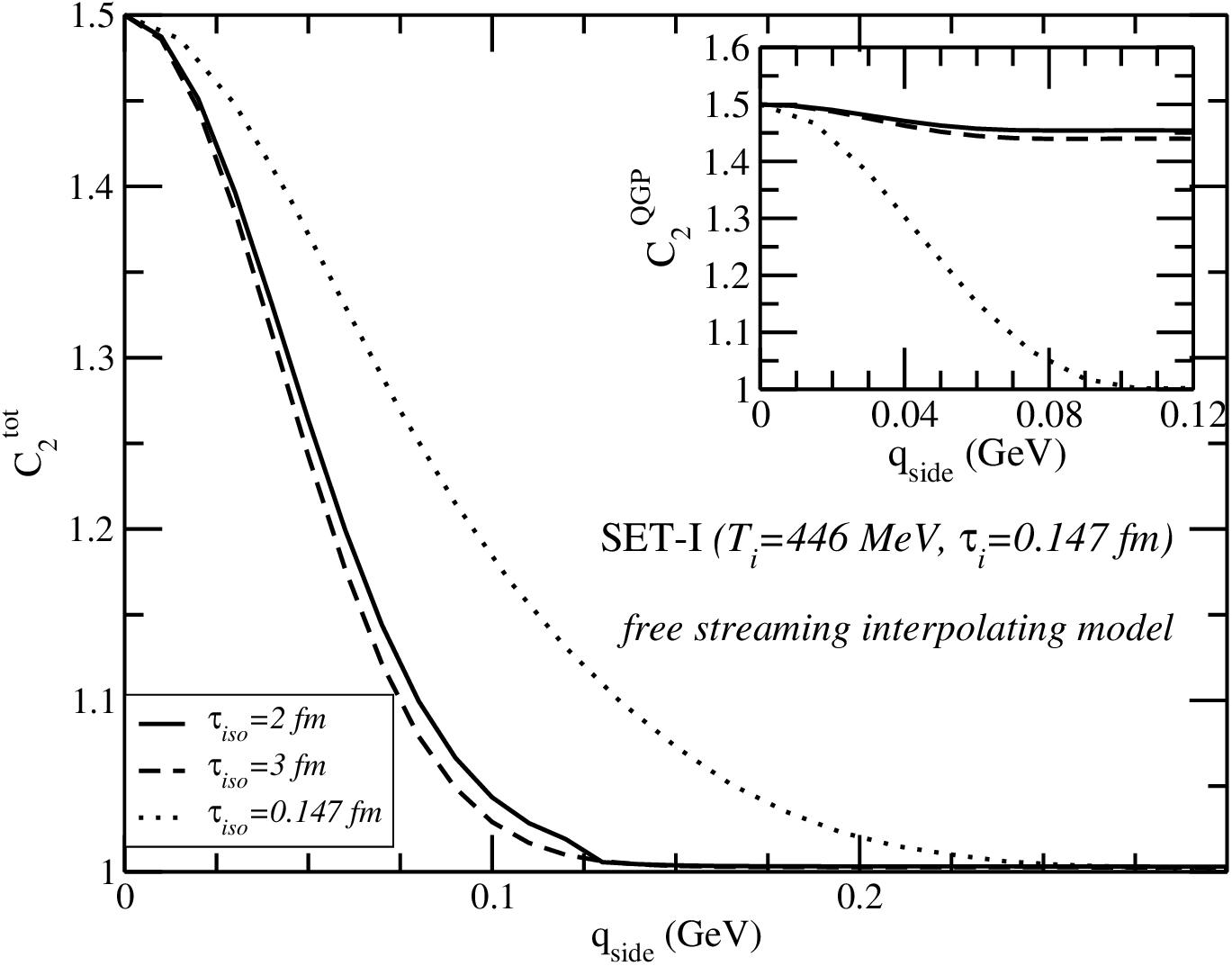}
\caption{Correlation function for photon pairs as a function of 
$q_{side}$  for SET-I ($T_i=446$ MeV and $\tau_i=0.147$ fm/c) 
is plotted with different $\tau_{iso}$ 
and the inset figure is same for QGP (aQGP+iQGP) phase only.}
\label{fig_c2_qside_446}
\end{center}
\eef
\bef
\begin{center}
\includegraphics[scale=0.3]{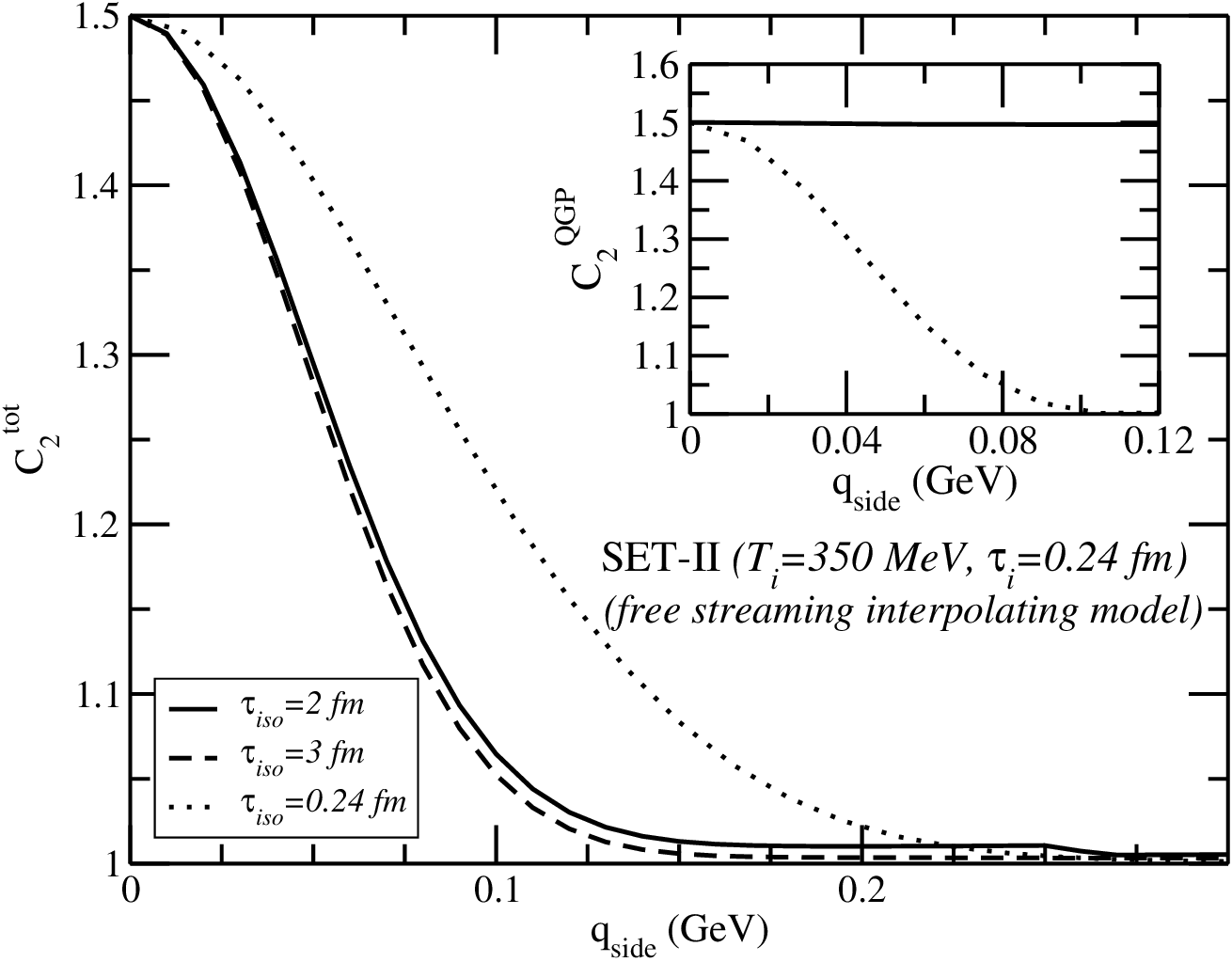}
\caption{Correlation function for photon pairs as a function 
of $q_{side}$  for SET-II ($T_i=350$ MeV and $\tau_i=0.24$ fm/c) 
is plotted with different $\tau_{iso}$
and the inset figure is same for QGP (aQGP+iQGP) phase only.}
\label{fig_c2_qside_350}
\end{center}
\eef
By  taking $k_{1T}$ = $k_{2T}=$ 2 GeV, $y_1=y_2$=0 and fixing 
$\psi_2$=0 and 
varying $\psi_1$, we obtain $C_2$ as a function of  $q_{side}$. 
 In Figs.~\ref{fig_c2_qside_446} and \ref{fig_c2_qside_350}, we have plotted the 
variation of $C_2$ as a function of $q_{side}$ for the full evolution scenario with  
SET-I and SET-II initial conditions respectively. 
From both the figures, it is observed that $C_2$ is shifted towards left considerably.
By increasing  $\tau_{iso}$, 
the value of $R_{side}$ (see table~\ref{tb_radii}) which corresponds to $q_{side}$ is also enhanced. 
This happens because by increasing $\tau_{iso}$, the system expands slower to 
achieve thermalization and isotropization. 
Whereas $C_2$ for the QGP (aQGP+isotropic QGP) phase  depicted in the insets of 
Figs.\ref{fig_c2_qside_446} and \ref{fig_c2_qside_350}  shows an opposite behaviour compared 
to that observed in the full evolution scenario. In other words, by increasing $\tau_{iso}$ the 
values of $R_{side}$  increase in full evolution scenario whereas it decrease in the QGP phase. 
It can be shown that $R_{side}\sim 1/(1+E_{\mathrm collective}/E_{\mathrm thermal})$~\cite{pratt}, where 
 $E_{thermal}$ depends inversely on $\tau_{iso}$ . In addition to it, 
the flow is not developed properly in the QGP phase, so $E_{collective}\ll E_{thermal}$. 
Thus, with the increase of  $\tau_{iso}$ the ratio   
$E_{\mathrm collective}/E_{\mathrm thermal} $ increases. As a result 
the value of $R_{side}$ decreases. Whereas in the hadronic phase, the flow is  fully developed 
resulting in  $E_{collective}\gg E_{thermal}$. The thermal energy is reduced even more by 
increasing $\tau_{iso}$. So due to the radial flow effect the  values of $R_{side}$ increase with 
the increase in the values of $\tau_{iso}$ in the full evolution scenario. 
\subsection{$C_2$ as function of $q_{long}$} 
\bef[!htbp]
\begin{center}
\includegraphics[scale=0.3]{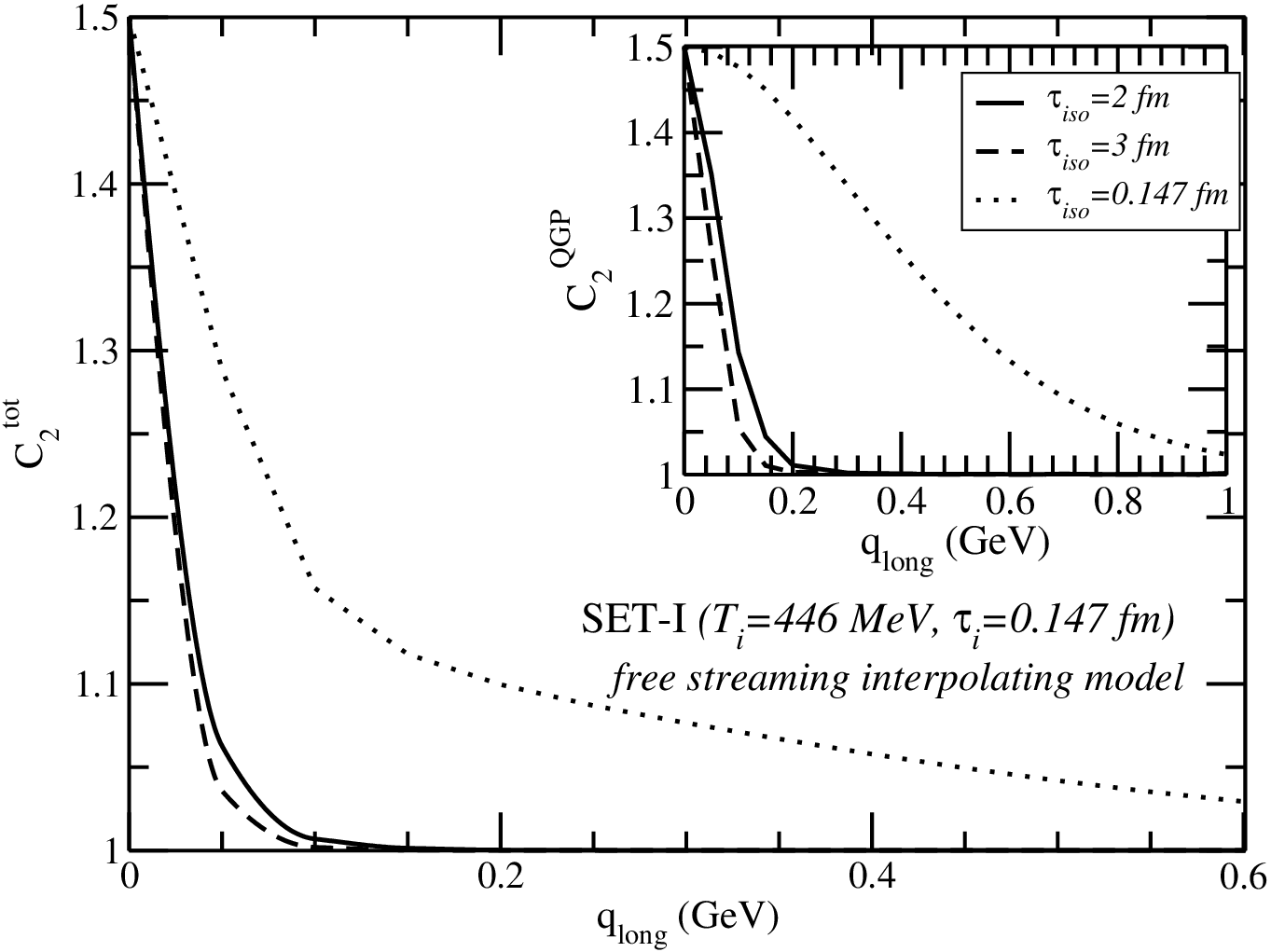}
\caption{Correlation function for photon pairs as a function of 
$q_{long}$  for $T_i=446$ MeV and $\tau_i=0.147$ fm/c
and the inset figure is same for QGP (aQGP+iQGP) phase only.}
\label{fig_c2_qlong_446}
\end{center}
\eef
\bef[!htbp]
\begin{center}
\includegraphics[scale=0.3]{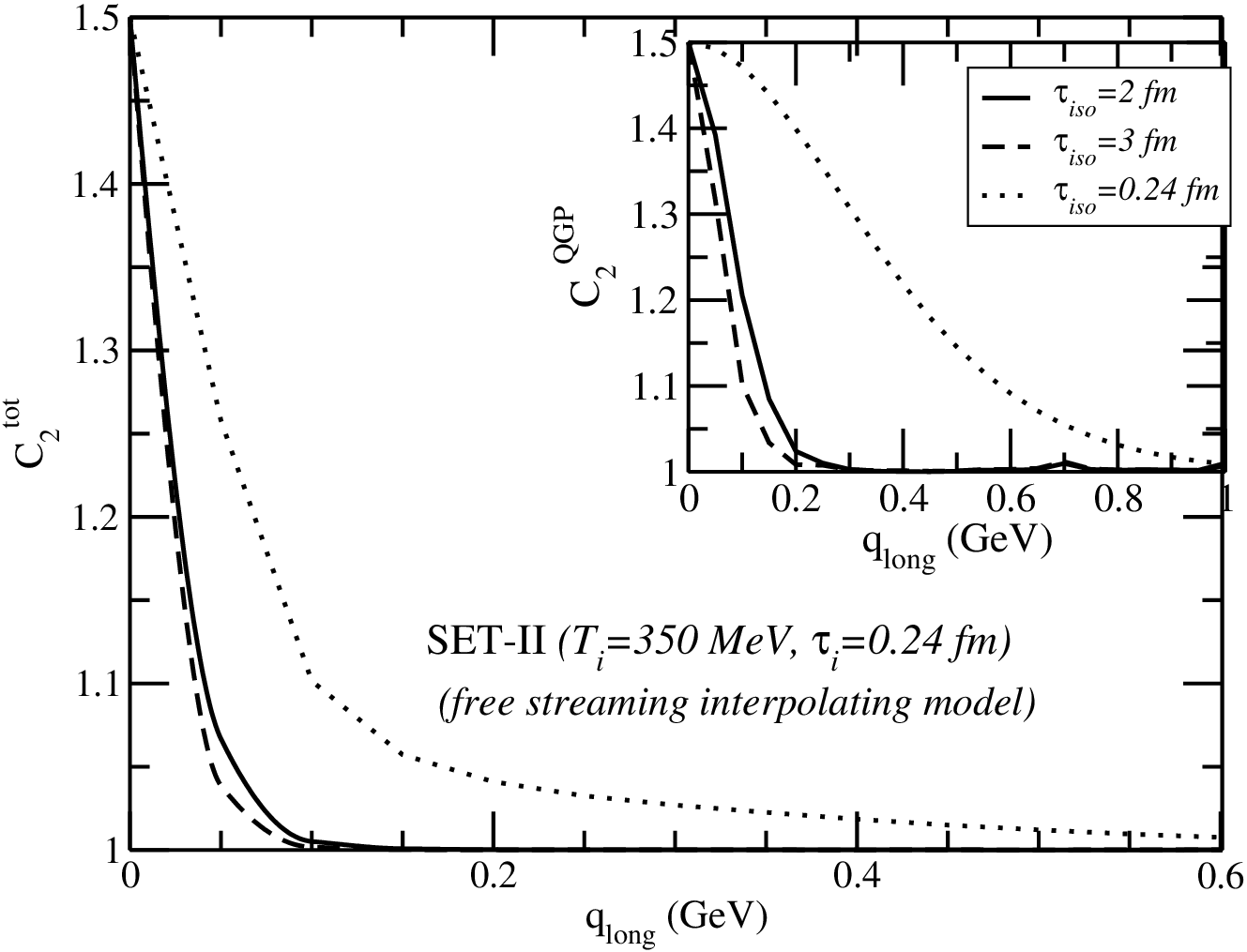}
\caption{Correlation function for photon pairs as a function of 
$q_{long}$  for $T_i=350$ MeV and $\tau_i=0.24$ fm/c
and the inset figure is same for QGP (aQGP+iQGP) phase only.}
\label{fig_c2_qlong_350}
\end{center}
\eef

$C_2$ as a function of  $q_{long}$ can be obtained by taking 
$\psi_1=\psi_2$=0, $k_{1T}=k_{2T}$=2 GeV and taking 
one of the photons at mid-rapidity ($y_{1}$ = 0) and 
varying the other ($y_{2}$). The variation of $C_2$ 
as a function of $q_{long}$ with  SET-I and SET-II initial 
conditions for RHIC energy is shown in 
Figs.~\ref{fig_c2_qlong_446} and \ref{fig_c2_qlong_350} respectively. 
It is clear from both the figures that there is a considerable 
difference for isotropic (when $\tau_{iso}=\tau_i$) and anisotropic 
(for $\tau_{iso}$= 2, 3 fm/c) scenarios.  
It is argued previously that the anisotropy in momentum space arises 
due to $\la p_L^2 \ra \ll \la p_T^2 \ra $. Thus  we can argue here 
that the difference arises in size in the longitudinal direction because of 
the above said asymmetry in momentum space. Hence  $R_{long}$  increases 
(see Table~\ref{tb_radii}) with the increase of $\tau_{iso}$. 

\subsection{Source Dimensions}

\begin{table}[!htbp]
\begin{center}
 \caption{The values of $R_{out}$, $R_{side}$ and $R_{long}$ obtained 
 from $C_2$ (using Eq.~\ref{eq_prmt})  is tabulated below.}
\label{tb_radii}
\vskip1.0cm
\begin{tabular}{|c|c|c|c|c|}
\hline
$T_i$ (MeV) &$\tau_{iso} (fm/c)$  & $R_{out}$ (fm)& $R_{side}$ (fm) & $R_{long}$ (fm) \\
\hline
446 & 0.147 &4.5 & 1.95& 2.6 \\
&2.0&5.5&3.13&6.6\\
&3.0&5.6&3.34&6.8\\
\hline
350 & 0.24 &4.69 &1.77&2.9 \\
& 2 &6.08& 2.83& 6.3\\
& 3 &6.09 & 2.96& 6.7\\
\hline

\end{tabular}
\end{center}
\end{table}

 We would like to mention here that the
HBT radii give the length of homogeneity of the source and
this is equal to the geometric size if the source is static. 
The HBT radii obtained from  $C_2$ using Eq.~\ref{eq_prmt} is 
tabulated in Table.~\ref{tb_radii}.
However, for a dynamic source, e.g., for the system formed after
ultra-relativistic heavy ion collisions, the HBT radii are
smaller than the geometric size. Figuring out the numbers tabulated in 
Table.~\ref{tb_radii}, it is clear that the HBT radii increase
with the inclusion of anisotropy, i.e., with increasing $\tau_{iso}$. 
Also a remarkable change in HBT radii is observed in the QGP phase 
(for both $R_{side}$ and $R_{long}$) with the inclusion of 
momentum space anisotropy except the outward direction.
\section{Summary} 
\label{sec_summ}
In this work, we have attempted to evaluate the correlation function, $C_2$ 
for two identical photons as functions of  $q_{out}$, $q_{side}$  and $q_{long}$ 
for RHIC energy with two sets of initial conditions and with initial state momentum 
space anisotropy. Hence $R_{out}$, $R_{side}$  and $R_{long}$ extracted from 
 $C_2$ in such a scenario  provide us the spatial informations of the 
evolving system. 
In summary, we have shown that $C_2$ for  QGP phase doesn't change appreciably 
with $q_{out}$ for both sets of initial conditions used here. For the entire 
evolution, we do observe an appreciable change in the variation of $C_2$ as a function of
$q_{out}$ due to the effect of radial flow and $\tau_{iso}$ dependent initial 
conditions for hydrodynamic evolution. We also observe a significant change  
for both the QGP phase and full evolution for the initial momentum space 
anisotropy when $C_2$ is plotted as a function of $q_{side}$.  The most interesting 
results are obtained in the variation of $C_2$ along the longitudinal direction because 
of the asymmetry in $p_T-p_L$ plane. As $\la p_L^2 \ra \ll \la p_T^2 \ra $, the maximal effect of momentum anisotropy 
is observed and correspondingly $R_{long}$ changes quite substantially with $\tau_{iso}$ 
for both sets of initial conditions. 
Hence, it is observed that by increasing the values of $\tau_{iso}$, 
all the HBT radii increases in full evolution scenario considerably due to 
effect of initial momentum anisotropy and the radial flow as well. 
However, in the QGP phase it affects only in side-ward and longitudinal directions. 
These are the most remarkable 
results that have been obtained in this work by introducing initial state momentum anisotropy. 
Finally, it is to be noted that we have not considered the fact that how much fractions 
of iQGP and aQGP are there during the transition from aQGP to iQGP at $\tau=\tau_{iso}$. 
Although, in principle, this aspect should be considered during this transition.  
This concept will be incorporated in the 
lepton-pair interferometry and work in this line is in progress~\cite{dilep_aniso}.
It is also straight forward to extend this analysis to LHC energies.
\section{ Acknowledgment:}
P M and M M want to thank L. Bhattacharya for useful discussion.

\normalsize

\end{document}